# Surfactant Antimony enhanced Indium incorporation on InGaN (000$\bar{1}$) surface: a DFT study

Yiou Zhang and Junyi Zhu[*]

*Department of physics, the Chinese University of Hong Kong, Shatin, N.T. 999077, Hong Kong*

InGaN is an ideal alloy system for optoelectronic devices due its tunable band gap. Yet high-quality InGaN requires high In concentration, which is a challenging issue that limits its use in green-light LEDs and other devices. In this paper, we investigated the surfactant effect of Sb on the In incorporation on InGaN (000$\bar{1}$) surface via first-principles approaches. Surface phase diagram was also constructed to determine surface structures under different growth conditions. By analyzing surface stress under different structures, we found that Sb adatom can induce tensile sites in the cation layer, enhancing the In incorporation. These findings may provide fundamental understandings and guidelines for the growth of InGaN with high In concentration.

Keywords: A1. Surfaces, B1. Nitrides, B2. Semiconducting III-V materials, A1. Computer simulation, B1. Gallium compounds.

## I. Introduction

In recent years, InGaN based optoelectronic devices have attracted great attention due to their superior material properties and flexible choices of the band gap [1-8]. A tunable direct band gap ranging from 0.7eV for InN [9-12] to 3.4eV for GaN [12-14] can be achieved by varying the concentration of In in InGaN [15,16], covering the whole visible light range. Therefore, InGaN alloys are ideal materials for the fabrication of white light-emitting-diodes (LEDs), laser-diodes (LDs), as well as high-efficiency solar-cells [17-19].

Despite the great success in InGaN-based blue-light LEDs [1,8,15,20,21], high-efficiency green-light LEDs with a high In concentration is still challenging, especially in organometallic vapor phase epitaxy (OMVPE). Due to the large bond length difference between GaN and InN, segregated In ad-layers or In rich islands on the surface are often observed in the OMVPE growth of InGaN [8,22-24]. Such large size mismatch may also lead to spinodal decomposition, resulting in a large miscibility gap and a limited In concentration in InGaN [14]. In addition to the thermodynamics, kinetics also has a strong influence on the In incorporation. Due to the low thermo-stability of InN [25], growth temperature of InGaN is usually lower than that of GaN, leading to limited surface kinetics, rough surface morphology, and poor crystal quality. [8] All these problems may largely reduce the efficiency of InGaN based devices with high In content, thereby limiting the use of InGaN-based green-light LEDs. [8,14,26]

Surface orientation and polarity also play important roles in the growth of InGaN. A typical growth direction of InGaN is the c-direction of wurtzite structure, where the surfaces belong to Tasker's type-III polar surfaces [27]. In addition to the c-plane, (0001) (In,Ga)-polar surface [28], the crystallographically inequivalent (000$\bar{1}$) N-polar surface, has also been found to be important recently [22]. Compared with In-polar InN, the dissociation temperature of N-polar InN is at most 100℃ higher

---

[*] Corresponding authors. Tel. +852 39436365

E-mail address: jyzhu@phy.cuhk.edu.hk (J. Zhu)

[29-33], and thus N-polar InGaN may be grown at a higher temperature with a better crystal quality. Also, enhanced In incorporation was observed on N-polar surface for both OMVPE [28] and plasma-assisted molecule beam epitaxy (MBE) [29,34]. First-principle calculations show that In atoms on N-polar surface of InGaN have a weaker surface segregation but higher thermal stability compared with that on (In,Ga)-polar surface [22]. Therefore, N-polar InGaN may be conducive to a higher In concentration and a better crystal quality.

Use of surfactants during epitaxial growth can be an appealing strategy for the modification of surface thermodynamics and surface kinetics [35-42]. Surfactants are elements that always flow on top of the growing front [40-44], which have been used in the epitaxial growth of III-V semiconductors [40-42,45-50]. By adding a small amount of Sb, Bi, or In as surfactants during GaN growth, the mobility of surface adatoms can be enhanced and the surface smoothness can be improved. [47-51] In addition to modifying surface kinetic processes, Bi and Sb surfactants have also been proved to be useful in changing the surface reconstruction and dopant concentrations in other III-V semiconductors growth [40,41,43,45,46]. Studies of surfactant effects during GaN growth only considered (0001) Ga-polar surface, while possible surfactant effects on (000$\bar{1}$) N-polar surface have not been thoroughly investigated. Studies of surfactant effects during InGaN growth are also limited. It was found that an improvement on crystal quality and optical performance can be achieved by adding some Sb during growth of InGaN. [52] Yet, to our knowledge, the structure properties of InGaN surface with Sb as surfactants have not been studied.

Recently, it was found that an abrupt increase in In concentrations was achieved by adding a small amount of Sb on InGaN surfaces during OMVPE growth [26]. Such enhanced In incorporation was observed only when Sb concentration reached a critical limit, which indicated that a phase change on the surface is possible [26]. The underlying physical mechanism is still unknown, because direct observation of the surface during OMVPE growth is very difficult, if not impossible. Also, it is hard to determine the polarity of InGaN from experiments. Owing to advantages of N-polar surface on In incorporation, first principles calculations of the surfactant assisted In incorporation on (000$\bar{1}$) surface can be appealing here.

Since InGaN is usually grown pseudomorphically on GaN, and the (000$\bar{1}$) surfaces of InGaN and GaN have high similarities, pure GaN can be used as a host system to avoid complexity on the random alloy in the modeling process. Still, to understand the surfactant enhanced In incorporation effect in GaN, it is essential to first study the reconstructions of bare GaN and In or Sb passivated GaN (000$\bar{1}$) surfaces. Bare GaN (000$\bar{1}$) surface forms various reconstructions under different chemical potential conditions. It changes from 2×2 Ga adatom reconstruction under N-rich condition to 1×1 Ga adlayer reconstruction under Ga-rich limit, with the increase of Ga/N ratio. [53] As for InGaN, it was found by experiments [23] and theoretical studies [54] that the (000$\bar{1}$) surface is covered by a metallic layer consisting of both In and Ga. Although additional In or Ga atoms on the ad-layer may further change the morphology and periodicity of surface [23], their effects on InGaN surface are similar. To our knowledge, Sb involved surface reconstructions on GaN (000$\bar{1}$) surface have not been studied. Nevertheless, the empirical electron-counting-rule (ECR) [55-57] shows a general applicability for surface reconstructions on III-V semiconductor surfaces [43,53,58-60]. Hence possible surface structures of Sb on GaN (000$\bar{1}$) surface can be searched and clarified based on ECR.

In this paper, we studied Sb surfactant's effects on In incorporation on (000$\bar{1}$) N-polar surface of InGaN through first-principle calculations. Various surface reconstructions involving Ga, N, In, and Sb were calculated. Based on the surface reconstruction results, we further calculated the formation energy of of In atoms occupying Ga sites at the surface layer. Our calculations indicate that Sb ad-atom reconstruction would induce significant charge redistributions among surface atoms, resulting in tensile sites on the first surface cation layer. As a consequence, the incorporation of In into GaN can be enhanced. In addition, our studies indicate that under certain concentration of Sb on the growth front, an increased partial pressure of In does not necessarily enhance In incorporation. These findings may provide important guidelines to grow high quality InGaN based green-light LEDs.

**II. Computational Approach**

Total energy calculations are based on Density Functional Theory [61,62] as implemented in VASP code [63,64], with a plane wave basis set [65,66]. The energy cutoff of the plane wave was set at 400eV, with d electrons of In and Ga included explicitly as valence electrons. PBE Generalized Gradient Approximation (GGA) functional [67] was used as the exchange-correlation functional. The $(000\bar{1})$ surface of GaN is modeled using a slab geometry with eight III-V bilayers, and a vacuum region of at least 20Å. (4×4) unit cell slabs were used with Γ-centered (2×2×1) Monkhorst-Pack mesh [68] for k-point sampling. Convergence tests with respect to k-point sampling, energy cutoff, slab and vacuum thickness were explicitly performed. Transition state energy was determined by the climbing image nudged elastic band (CI-NEB) method [81]. For all the slabs, pseudo-hydrogen atoms of fractional charge $q = 1.25e$ were used to passivate the dangling bonds of bottom Ga atoms. All the atoms were relaxed until the force converged to less than $0.01 eV/Å$. Calculations of In incorporation into the first bilayer on the surface were based on two cases: one $In_{Ga}$ in each (4×4) cell (6.25% In concentration in the first bilayer), and one $In_{Ga}$ in each (2×2) cell (25% In concentration in the first bilayer), corresponding to low In content and moderate In content respectively, which are reasonably close to the In concentration in experiments [6,15-17,21,26].

Due to the low symmetry of wurtzite structures, it is difficult to calculate the absolute surface energies of polar surfaces [69]. Nevertheless, the relative stabilities of different surface reconstructions can be determined from relative surface energies with respect to the relaxed unreconstructed surfaces, through the following equation:

$$\Delta\gamma(\mu_{Ga},\mu_N,\mu_{In},\mu_{Sb}) = E_S^{tot} - E_{bare}^{tot} - \Delta n_{Ga}\mu_{Ga} - \Delta n_N\mu_N - \Delta n_{In}\mu_{In} - \Delta n_{Sb}\mu_{Sb}, (1)$$

where $E_S^{tot}$ and $E_{bare}^{tot}$ are the total energies of the considered surface and the bare surface, respectively. $\Delta n_X$ represents the difference of the number of X (X = Ga, N, In, Sb) atoms between the considered surface and the bare surface, and $\mu_X$ is the corresponding chemical potential of X element. For calculations of In incorporation, the (average) formation enthalpy of $In_{Ga}$ is determined by the formula:

$$\Delta H(In_{Ga}) = \frac{E_{In-doped}^{tot} - E_{undoped}^{tot} - n(In_{Ga})(\mu_{Ga}-\mu_{In})}{n(In_{Ga})}, (2)$$

where $E_{In-doped}^{tot}$ and $E_{undoped}^{tot}$ are total energies of slabs with or without In incorporated into the first bi-layer, and both slabs are under the same surface reconstruction. $n(In_{Ga})$ is the number of $In_{Ga}$ substitution sites in the slab. When the surface is at thermal equilibrium, it is required that

$$\mu_{Ga} + \mu_N = \mu_{GaN}, (3)$$

where $\mu_{GaN}$ is the total energy of bulk GaN per formula unit. The upper limit of $\mu_X$ corresponds to X-rich limit, and is determined by the total energy per atom of bulk Ga, one nitrogen molecule, bulk In, and bulk Sb:

$$\Delta\mu_X = \mu_X - \mu_X^{bulk} \leq 0. \quad (4)$$

In addition, all the competing secondary phases should be thermodynamically unstable:

$$\mu_{In} + \mu_N \leq \mu_{InN}, \quad (5)$$
$$\mu_{Ga} + \mu_{Sb} \leq \mu_{GaSb}, \quad (6)$$
$$\mu_{In} + \mu_{Sb} \leq \mu_{InSb}. \quad (7)$$

Since previous theoretical works [49,50] suggested that SbN can easily form on Ga-polar (0001) surface of GaN, we have also considered pure phase of SbN. Yet our calculations show that such phase is thermodynamically unstable, since the calculated formation enthalpy is larger than zero. This indicates that bonding between Sb and N is likely to be favored only under certain surface electronic environment. Throughout the calculations, the N-rich limit was considered, as large V/III ratio can also improve In incorporation [8]. Structure parameters of all the considered elementary solids and secondary phases are within 2% differences with experimental data [70-75]. In particular, the calculated lattice parameters of GaN are within 0.5% difference with experimental result [76], and the calculated formation enthalpy (-1.00eV) is close to experimental value (-1.17eV [77]). Although the calculated formation enthalpy of InN was positive due to error of GGA functional [78], such error will only influence the chemical potential of In under In-rich limit, and our conclusions will not be affected.

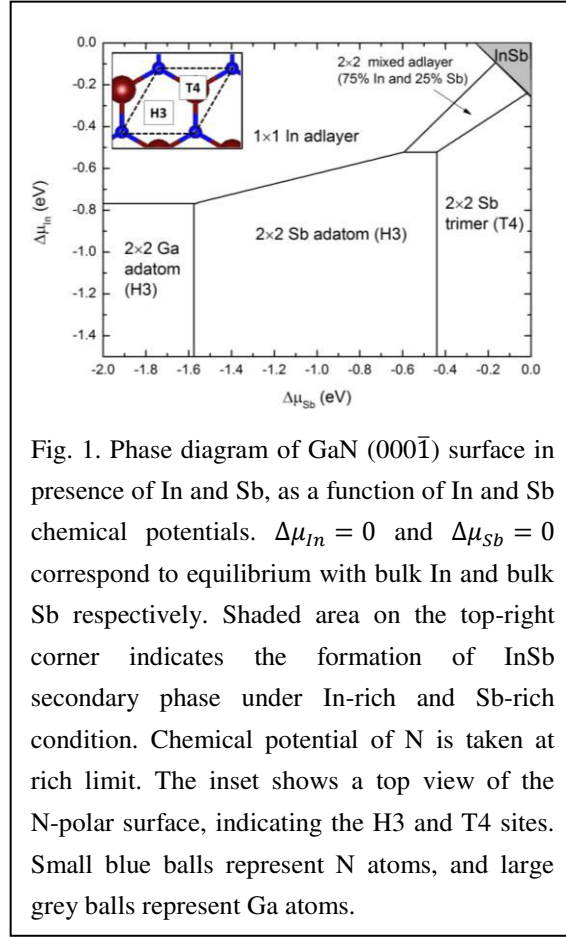

Fig. 1. Phase diagram of GaN $(000\bar{1})$ surface in presence of In and Sb, as a function of In and Sb chemical potentials. $\Delta\mu_{In} = 0$ and $\Delta\mu_{Sb} = 0$ correspond to equilibrium with bulk In and bulk Sb respectively. Shaded area on the top-right corner indicates the formation of InSb secondary phase under In-rich and Sb-rich condition. Chemical potential of N is taken at rich limit. The inset shows a top view of the N-polar surface, indicating the H3 and T4 sites. Small blue balls represent N atoms, and large grey balls represent Ga atoms.

### III. Results and Discussions

To model different surface structures and geometries under different In and Sb concentration, various surface reconstructions have been considered. Surface reconstructions involving Ga, N, and In were searched based on previous literatures [23,53,54]. Multi-layers of In on the surface were not considered since they become energetically favorable only near In-rich limit, and their effects on the GaN bilayers are similar to that on one ad-layer of In. Surface reconstructions of Sb were searched based on the empirical ECR [55-57]. Since each N dangling bond contains 5/4 electrons, it is required that 3 electrons should be transferred to N dangling bonds on a 2×2 cell. The 2×2 Sb-adatom model and the 2×2 Sb-trimer model were considered, both of which can be placed on H3 site (above one sub-layer hollow site) and T4 site (above one sub-layer Ga atom), as shown in the inset of Fig. 1. We have also considered Sb substituting Ga or N in the first bi-layer, yet the resulting surface energies of these configurations are always higher than Sb-adatom or Sb-trimer model. In addition, a mixed Sb/In ad-layer with different Sb concentrations was considered. The surface phase diagram is shown in Fig. 1. Under In-poor and Sb-poor condition, GaN $(000\bar{1})$ surface is terminated by Ga adatom. With the increase of Sb concentration, the most stable surface reconstruction changes to an Sb adatom structure and then to an Sb trimer structure. Under In-rich condition, the surface is always covered by an In ad-layer, consistent with previous experimental and theoretical studies [22,23,54]. Since H is usually present in OMVPE growth, it is also essential to consider its effect on the surface structures. However, it is largely determined by the growth environment [41]. Here, we can estimate the rich limit of H. The limit can be set as $\mu_H = \frac{1}{2}E_{H_2}^{tot} - 0.75eV$, where $E_{H_2}^{tot}$ is the total energy $H_2$ molecule at zero temperature from DFT calculations, and -0.75eV is the entropy contribution to free energy at 1000K and 1atm, including zero-point energy [80]. Under such H-rich limit, H-covered surface is thermodynamically more stable than Sb adatom or trimer structure, even under Sb-rich condition. Nevertheless, Sb adatom or trimer structure may still be thermodynamically stable under experimental condition, as the concentration of atomic H should be lower. Also, even under this H-rich condition, adsorption energy of H on Sb adatom structure is at least 0.45eV, much higher than -1.92eV on bare GaN surface, which indicates that the adsorption of H is largely suppressed on Sb-covered surface. These results indicate that there might be a competition between Sb and H coverage on the surface, and Sb coverage may become more dominant under higher growth temperature or lower partial pressure of $H_2$ gas. Nevertheless, the growth temperature cannot be too high, as InGaN may become unstable due to In loss at high temperature.

Electronic structures of the surface reconstructions can be important to understand the mechanism of In incorporation into the first cation layer near the top surface. For In or InSb ad-layer reconstructions, as shown in Fig. 2(a), each N atom on the surface are bonded to one In or Sb atom in the ad-layer, thus the abundant electrons from the metallic ad-layer will passivate the N dangling bonds. As a consequence, GaN bilayer on the surface has a similar structure and stress distribution with bilayers in bulk GaN. On the other hand, for adatom or trimer reconstructions, as shown in Fig. 2(b) and Fig. 2(c), in each 2×2 cell, three of the surface N atoms are 4-fold bonded, while one N atom (resatom) remains 3-fold bonded. This N resatom will attract extra electrons from the adatom or the trimer, thus adopting $sp^2$-like configuration with lone-pair electrons in the $p_z$-orbital. Such changes in electronic structures will further affect the electronic structures on the surface. As shown in Fig. 2(d), three Ga atoms bonded to the N resatom are pulled towards the resatom, due to a shorter covalent radius of the $sp^2$-like N atom. The stress is further distributed among other N atoms. As a result, one Ga atom will be on a tensile site, as N atoms around this Ga atom relax away from the Ga atom. For H3 adatoms or trimers, such relaxations also lead to shorter bonds between surface N atoms and the adatoms or trimers, further stabilizing the structures. On the other hand, for T4 adatoms or trimers, such relaxations would slightly destabilize the surface adatoms or trimers. Therefore, stress effect is more significant for adatoms or trimers on H3 sites than those on T4 sites.

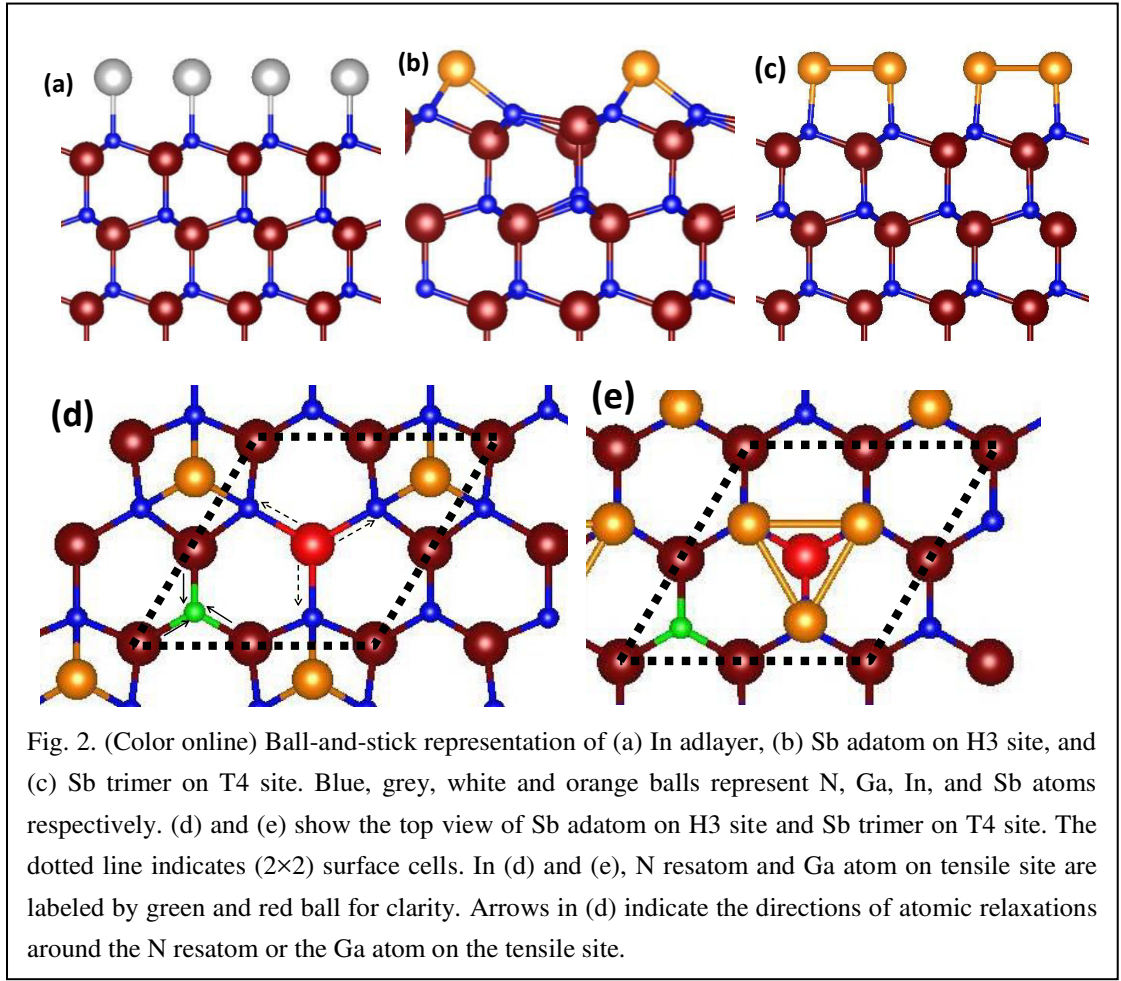

Fig. 2. (Color online) Ball-and-stick representation of (a) In adlayer, (b) Sb adatom on H3 site, and (c) Sb trimer on T4 site. Blue, grey, white and orange balls represent N, Ga, In, and Sb atoms respectively. (d) and (e) show the top view of Sb adatom on H3 site and Sb trimer on T4 site. The dotted line indicates (2×2) surface cells. In (d) and (e), N resatom and Ga atom on tensile site are labeled by green and red ball for clarity. Arrows in (d) indicate the directions of atomic relaxations around the N resatom or the Ga atom on the tensile site.

Since the radius of In is much larger than that of Ga, the tensile site on the surface may enhance In incorporation into the surface bilayer. As one tensile site is created in the first bi-layer of each 2×2 cell, In content may be increased in the first bilayer. Calculated formation enthalpies of $In_{Ga}$ under different surface reconstructions are shown in Fig. 3. The results indicate that both Ga adatom (H3) and Sb adatom (H3) can enhance In incorporation, while Sb trimer (T4) has little effect on the formation enthalpy. Compared to the Ga adatom, the Sb adatom has a larger stress effect, and remains stable under higher In concentration. Hence, compared with intrinsic surface reconstructions, surfactant Sb may enhance In incorporation into the first bilayer. In fact, as can be seen from Fig. 3(a), under suitable Sb and In concentrations, an $In_xGa_{1-x}N$ bilayer (with x=6.25%) on surface becomes energetically more favorable than the GaN bilayer, even without considering the mixing entropy. When the In concentration increases, the formation enthalpy of $In_{Ga}$ is increased. This is due to a stress compensation effect between $In_{Ga}$ sites. In both cases, the formation enthalpies of $In_{Ga}$ with presence of Sb can be lowered by 0.5eV, compared to the optimal growth condition without Sb. This mechanism shows a good consistency with pervious experimental findings, in which enhanced In incorporation was only observed when Sb concentration reached a critical value [26]. When Sb concentration is further increased, the surface may change from an Sb adatom reconstruction to an Sb trimer reconstruction. Nevertheless, Sb trimer or In adlayer do not have significant effects on In incorporation, thus enhancement on In incorporation can only happen under moderate In concentration and a certain Sb concentration. These results indicate that enhanced In content can be achieved by utilizing the surfactant effect of an optimal amount of Sb and increased partial pressure of In precursor is ineffective.

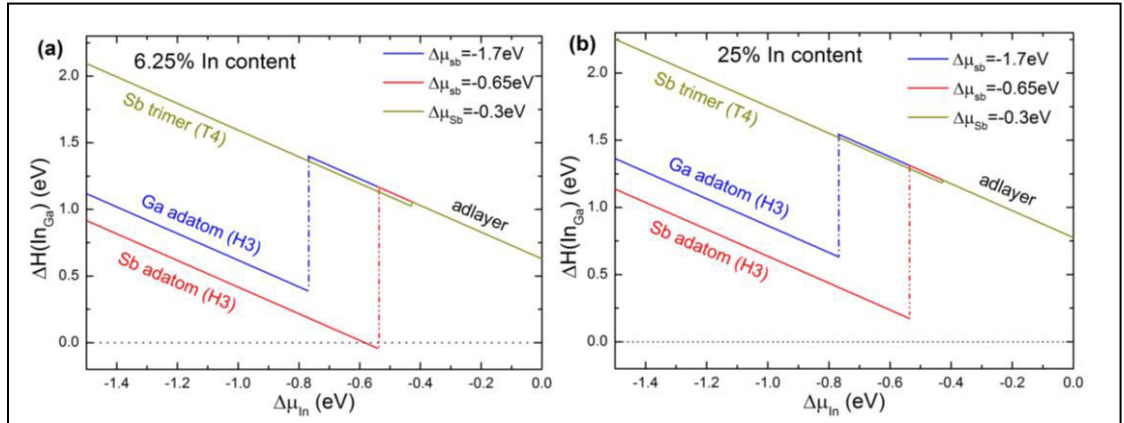

Fig. 3. (Color online) Formation enthalpies of In$_{Ga}$ in first bilayer of GaN as functions of In chemical potentials under different surface reconstructions, with (a) one In$_{Ga}$ in each 4×4 cell, and (b) one In$_{Ga}$ in each 2×2 cell. Different concentration of Sb has been considered: (1) Sb-poor ($\Delta\mu_{Sb}$=-1.7eV), (2) Sb-moderate ($\Delta\mu_{Sb}$=-0.65eV), and (3) Sb-rich ($\Delta\mu_{Sb}$=-0.3eV). Surface reconstructions are labeled beside the curve (In ad-layer and In, Sb-mixed ad-layer have similar effects on GaN bilayer, thus only labeled as ad-layer).

We also consider the kinetic barrier of In incorporation, by calculating diffusion barrier of In into the sub-layer by kick-out mechanism. The kinetic behaviors under both Sb adatom (H3) reconstruction and In adlayer reconstruction have been considered. For Sb adatom (H3) structure, one additional In atom is put on the surface, while for In adlayer structure one In atom in the adlayer diffuses into the cation layer. The calculated diffusion barrier is shown in Fig.4. As can be seen from the figure, diffusion barrier of In into the sub-surface cation layer is 1.70eV under Sb adatom (H3) reconstruction, much lower than that under In adlayer reconstruction (5.41eV). Such huge difference may be explained by the surface stress induced by Sb adatom, as the Ga atom on the tensile site is loosely bonded and easier to diffuse out. Also, energy gain from the shorter Sb-N bond length at the transition state configuration may give rise to lower transition state energy. Moreover, because surface segregation of In on (000$\bar{1}$) surface is not as likely as that on (0001) surface [22], and the energy barrier of cation inter-diffusion in bulk InGaN is high [79], the incorporated In is unlikely to diffuse outward during further growth. Therefore, In content in the first bilayer may be mostly preserved, enhancing the In concentration in bulk InGaN. Surfactant effects based on random InGaN alloys are also very interesting topics, yet out of the scope of this paper.

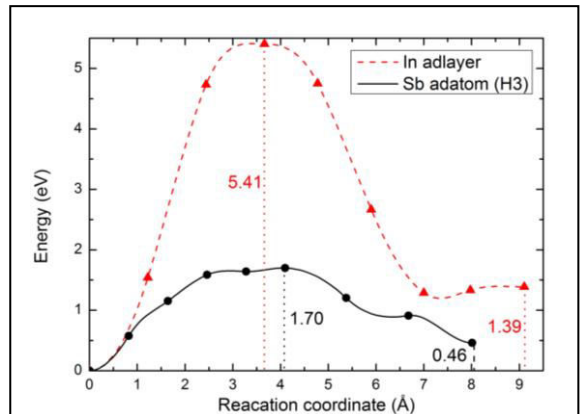

Fig.4. Diffusion barriers of In incorporation into the sub-layer under Sb adatom (H3) structure (black solid line) and In adlayer structure (red dashed line). For either path, energy of the initial structure is set as zero. Energies of the transition state and final state relative to the initial state are labeled on the figure.

## IV. Conclusion

In summary, by using GaN as the host system, we have studied the surfactant effect of Sb on InGaN (000$\bar{1}$) surface. Surface phase diagram shows that at moderate In and Sb concentrations, Sb adatom (H3) becomes energetically favorable. The surface reconstructions will change to Sb trimer or In ad-layer if the concentration of Sb or In is increased. Analysis on surface stress under different surface reconstructions indicates that adatoms on H3 sites lead to tensile sites in GaN bilayer, which can enhance In incorporation. Calculations of the formation enthalpies of $In_{Ga}$ under different In concentrations and different surface reconstructions show that Sb adatom structure can lower the formation enthalpy of $In_{Ga}$ by 0.5eV, and In concentration in thermodynamically stable InGaN may be increased. Our findings may provide basic physical understandings and fundamental guidelines for the growth of InGaN with high crystal quality as well as high In content, enlightening the fabrication of high-efficiency green-light LEDs.

**Acknowledgement**
This work was supported by the start-up funding and direct grant with the Project code of 4053134 at CUHK.

**References**
[1] S. Nakamura, The roles of structural imperfections in InGaN-based blue light-emitting diodes and laser diodes, Science. 281 (1998) 956-961.
[2] S. Nakamura, S. Pearton, G. Fasol, The blue laser diode: the complete story, Springer Science & Business Media, 2013.
[3] J. Northrup, L. Romano, J. Neugebauer, Surface energetics, pit formation, and chemical ordering in InGaN alloys, Appl. Phys. Lett. 74 (1999) 2319.
[4] Y. Nanishi, Y. Saito, T. Yamaguchi, M. Hori, F. Matsuda, T. Araki, A. Suzuki, T. Miyajima, MBE- growth, characterization and properties of InN and InGaN, physica status solidi (a). 200 (2003) 202-208.
[5] J.E. Northrup, GaN and InGaN (112_2) surfaces: Group-III adlayers and indium incorporation, Appl. Phys. Lett. 95 (2009) 133107.
[6] C. Skierbiszewski, Z. Wasilewski, I. Grzegory, S. Porowski, Nitride-based laser diodes by plasma-assisted MBE—From violet to green emission, J. Cryst. Growth. 311 (2009) 1632-1639.
[7] G. Stringfellow, Microstructures produced during the epitaxial growth of InGaN alloys, J. Cryst. Growth. 312 (2010) 735-749.
[8] F. Yam, Z. Hassan, InGaN: An overview of the growth kinetics, physical properties and emission mechanisms, Superlattices and Microstructures. 43 (2008) 1-23.
[9] P. Carrier, S. Wei, Theoretical study of the band-gap anomaly of InN, J. Appl. Phys. 97 (2005) 033707.
[10] J. Wu, W. Walukiewicz, K. Yu, J. Ager Iii, E. Haller, H. Lu, W.J. Schaff, Y. Saito, Y. Nanishi, Unusual properties of the fundamental band gap of InN, Appl. Phys. Lett. 80 (2002) 3967-3969.
[11] T. Matsuoka, H. Okamoto, M. Nakao, H. Harima, E. Kurimoto, Optical bandgap energy of wurtzite InN, Appl. Phys. Lett. 81 (2002) 1246-1248.
[12] P. Rinke, A. Qteish, M. Winkelnkemper, D. Bimberg, J. Neugebauer, M. Scheffler, Band gap and band parameters of InN and GaN from quasiparticle energy calculations based on exact-exchange density-functional theory, arXiv preprint cond-mat/0610141 (2006).
[13] E. Oh, S. Lee, S. Park, K. Lee, I. Song, J. Han, Optical properties of GaN grown by hydride vapor-phase epitaxy, Appl. Phys. Lett. 78 (2001) 273-275.

[14] I. Ho, G. Stringfellow, Solid phase immiscibility in GaInN, Appl. Phys. Lett. 69 (1996) 2701-2703.
[15] F. Ponce, D. Bour, Nitride-based semiconductors for blue and green light-emitting devices, Nature (1997).
[16] S. Nakamura, M. Senoh, S. Nagahama, N. Iwasa, T. Yamada, T. Matsushita, H. Kiyoku, Y. Sugimoto, InGaN-based multi-quantum-well-structure laser diodes, Japanese Journal of Applied Physics. 35 (1996) L74.
[17] C.J. Neufeld, N.G. Toledo, S.C. Cruz, M. Iza, S.P. DenBaars, U.K. Mishra, High quantum efficiency InGaN/GaN solar cells with 2.95 eV band gap, Appl. Phys. Lett. 93 (2008) 3502.
[18] O. Jani, I. Ferguson, C. Honsberg, S. Kurtz, Design and characterization of GaN∕ InGaN solar cells, Appl. Phys. Lett. 91 (2007) 132117.
[19] J. Wu, W. Walukiewicz, K. Yu, W. Shan, J. Ager Iii, E. Haller, H. Lu, W.J. Schaff, W. Metzger, S. Kurtz, Superior radiation resistance of $In_{1-x}Ga_xN$ alloys: Full-solar-spectrum photovoltaic material system, J. Appl. Phys. 94 (2003) 6477-6482.
[20] T. Mukai, M. Yamada, S. Nakamura, Characteristics of InGaN-based UV/blue/green/amber/red light-emitting diodes, Japanese Journal of Applied Physics. 38 (1999) 3976.
[21] S. Nakamura, Blue-green light-emitting diodes and violet laser diodes, MRS Bull. 22 (1997) 29-35.
[22] A.I. Duff, L. Lymperakis, J. Neugebauer, Understanding and controlling indium incorporation and surface segregation on $In_x Ga_{1-x} N$ surfaces: An ab initio approach, Physical Review B. 89 (2014) 085307.
[23] R.M. Feenstra, H. Chen, V. Ramachandran, A. Smith, D.W. Greve, Reconstructions of GaN and InGaN surfaces, Appl. Surf. Sci. 166 (2000) 165-172.
[24] S.Y. Karpov, R. Talalaev, I.Y. Evstratov, Y.N. Makarov, Indium Segregation Kinetics in MOVPE of InGaN‐ Based Heterostructures, physica status solidi (a). 192 (2002) 417-423.
[25] C. Gallinat, G. Koblmüller, J. Brown, J. Speck, A growth diagram for plasma-assisted molecular beam epitaxy of In-face InN, J. Appl. Phys. 102 (2007) 064907.
[26] J.L. Merrell, F. Liu, G.B. Stringfellow, Effect of surfactant Sb on In incorporation and thin film morphology of InGaN layers grown by organometallic vapor phase epitaxy, J. Cryst. Growth. 375 (2013) 90-94.
[27] P. Tasker, The stability of ionic crystal surfaces, Journal of Physics C: Solid State Physics. 12 (1979) 4977.
[28] S. Keller, N. Fichtenbaum, M. Furukawa, J. Speck, S. DenBaars, U. Mishra, Growth and characterization of N-polar InGaN/GaN multiquantum wells, Appl. Phys. Lett. 90 (2007) 1908.
[29] D.N. Nath, E. Gür, S.A. Ringel, S. Rajan, Molecular beam epitaxy of N-polar InGaN, Appl. Phys. Lett. 97 (2010) 071903.
[30] K. Xu, A. Yoshikawa, Effects of film polarities on InN growth by molecular-beam epitaxy, Appl. Phys. Lett. 83 (2003) 251-253.
[31] G. Koblmüller, C. Gallinat, S. Bernardis, J. Speck, G. Chern, E. Readinger, H. Shen, M. Wraback, Optimization of the surface and structural quality of N-face InN grown by molecular beam epitaxy, Appl. Phys. Lett. 89 (2006) 071902.
[32] G. Koblmüller, C. Gallinat, J. Speck, Surface kinetics and thermal instability of N-face InN grown by plasma-assisted molecular beam epitaxy, J. Appl. Phys. 101 (2007) 083516.

[33] H. Naoi, F. Matsuda, T. Araki, A. Suzuki, Y. Nanishi, The effect of substrate polarity on the growth of InN by RF-MBE, J. Cryst. Growth. 269 (2004) 155-161.

[34] D.N. Nath, E. Gür, S.A. Ringel, S. Rajan, Growth model for plasma-assisted molecular beam epitaxy of N-polar and Ga-polar $In_xGa_{1-x}N$, Journal of Vacuum Science & Technology B: Microelectronics and Nanometer Structures. 29 (2011) 021206.

[35] C. Herring, Some theorems on the free energies of crystal surfaces, Physical Review. 82 (1951) 87.

[36] P. Hartman, W. Perdok, On the relations between structure and morphology of crystals. I, Acta Crystallogr. 8 (1955) 49-52.

[37] P. Hartman, W. Perdok, On the relations between structure and morphology of crystals. II, Acta Crystallogr. 8 (1955) 521-524.

[38] P. Hartman, W. Perdok, On the relations between structure and morphology of crystals. III, Acta Crystallogr. 8 (1955) 525-529.

[39] W. Chou, C. Kuo, H. Cheng, Y. Chen, F. Tang, F. Yang, D. Shu, C. Liao, Effect of surface free energy in gate dielectric in pentacene thin-film transistors, Appl. Phys. Lett. 89 (2006) 112126.

[40] J. Zhu, G. Stringfellow, F. Liu, Dual-Surfactant effect on enhancing different p-type doping in GaP. 1 (2009) 21012.

[41] J. Zhu, F. Liu, G. Stringfellow, Enhanced cation-substituted p-type doping in GaP from dual surfactant effects, J. Cryst. Growth. 312 (2010) 174-179.

[42] J. Zhu, S. Wei, Overcoming doping bottleneck by using surfactant and strain, Frontiers of Materials Science. 5 (2011) 335-341.

[43] J. Zhu, F. Liu, G. Stringfellow, Dual-surfactant effect to enhance p-type doping in III-V semiconductor thin films, Phys. Rev. Lett. 101 (2008) 196103.

[44] M. Copel, M. Reuter, E. Kaxiras, R. Tromp, Surfactants in epitaxial growth, Phys. Rev. Lett. 63 (1989) 632.

[45] C. Fetzer, R. Lee, J. Shurtleff, G. Stringfellow, S. Lee, T. Seong, The use of a surfactant (Sb) to induce triple period ordering in GaInP, Appl. Phys. Lett. 76 (2000) 1440-1442.

[46] M. Pillai, S. Kim, S. Ho, S. Barnett, Growth of $In_xGa_{1-x}As$/GaAs heterostructures using Bi as a surfactant, J.Vac.Sci.Technol.B. 18 (2000) 1232-1236.

[47] L. Zhang, H. Tang, J. Schieke, M. Mavrikakis, T. Kuech, Influence of Bi impurity as a surfactant during the growth of GaN by metalorganic vapor phase epitaxy, J. Cryst. Growth. 242 (2002) 302-308.

[48] L. Zhang, H. Tang, J. Schieke, M. Mavrikakis, T.F. Kuech, The addition of Sb as a surfactant to GaN growth by metal organic vapor phase epitaxy, J. Appl. Phys. 92 (2002) 2304-2309.

[49] A.A. Gokhale, T.F. Kuech, M. Mavrikakis, A theoretical comparative study of the surfactant effect of Sb and Bi on GaN growth, J. Cryst. Growth. 303 (2007) 493-499.

[50] A.A. Gokhale, T.F. Kuech, M. Mavrikakis, Surfactant effect of Sb on GaN growth, J. Cryst. Growth. 285 (2005) 146-155.

[51] J.E. Northrup, Van de Walle, Chris G, Indium versus hydrogen-terminated GaN (0001) surfaces: Surfactant effect of indium in a chemical vapor deposition environment, Appl. Phys. Lett. 84 (2004) 4322-4324.

[52] K.G. Sadasivam, J. Shim, J.K. Lee, Antimony surfactant effect on green emission InGaN/GaN multi quantum wells grown by MOCVD, Journal of nanoscience and nanotechnology. 11 (2011) 1787-1790.

[53] D. Segev, Van de Walle, Chris G, Surface reconstructions on InN and GaN polar and nonpolar surfaces, Surf. Sci. 601 (2007) L15-L18.

[54] C.K. Gan, D.J. Srolovitz, First-principles study of In, Ga, and N adsorption on $In_xGa_{1-x}N$ (0001) and (000$\bar{1}$) surfaces, Physical Review B. 77 (2008) 205324.

[55] M. Pashley, Electron counting model and its application to island structures on molecular-beam epitaxy grown GaAs (001) and ZnSe (001), Physical Review B. 40 (1989) 10481.

[56] D. Chadi, Atomic structure of GaAs (100)‐(2×1) and (2×4) reconstructed surfaces, Journal of Vacuum Science & Technology A. 5 (1987) 834-837.

[57] W.A. Harrison, Theory of polar semiconductor surfaces, Journal of Vacuum Science and Technology. 16 (1979) 1492-1496.

[58] V. Ramachandran, C. Lee, R.M. Feenstra, A. Smith, J. Northrup, D.W. Greve, Structure of clean and arsenic-covered GaN (0001) surfaces, J. Cryst. Growth. 209 (2000) 355-363.

[59] N. Moll, A. Kley, E. Pehlke, M. Scheffler, GaAs equilibrium crystal shape from first principles, Physical Review B. 54 (1996) 8844.

[60] Van de Walle, Chris G, J. Neugebauer, Role of hydrogen in surface reconstructions and growth of GaN, Journal of Vacuum Science & Technology B. 20 (2002) 1640-1646.

[61] P. Hohenberg, W. Kohn, Inhomogeneous electron gas, Physical review. 136 (1964) B864.

[62] W. Kohn, L.J. Sham, Self-consistent equations including exchange and correlation effects, Physical Review. 140 (1965) A1133.

[63] G. Kresse, J. Hafner, Ab initio molecular-dynamics simulation of the liquid-metal–amorphous-semiconductor transition in germanium, Physical Review B. 49 (1994) 14251.

[64] G. Kresse, J. Furthmüller, Efficiency of ab-initio total energy calculations for metals and semiconductors using a plane-wave basis set, Computational Materials Science. 6 (1996) 15-50.

[65] P.E. Blöchl, Projector augmented-wave method, Physical Review B. 50 (1994) 17953.

[66] G. Kresse, D. Joubert, From ultrasoft pseudopotentials to the projector augmented-wave method, Physical Review B. 59 (1999) 1758.

[67] J.P. Perdew, K. Burke, M. Ernzerhof, Generalized gradient approximation made simple, Phys. Rev. Lett. 77 (1996) 3865.

[68] H.J. Monkhorst, J.D. Pack, Special points for Brillouin-zone integrations, Physical Review B. 13 (1976) 5188.

[69] C.E. Dreyer, A. Janotti, Van de Walle, Chris G, Absolute surface energies of polar and nonpolar planes of GaN, Physical Review B. 89 (2014) 081305.

[70] C. Yeh, Z. Lu, S. Froyen, A. Zunger, Zinc-blende–wurtzite polytypism in semiconductors, Physical Review B. 46 (1992) 10086.

[71] H.E. Swanson, E. Tatge, R.K. Fuyat, Standard X-ray diffraction powder patterns (1953).

[72] B.D. Sharma, J. Donohue, A refinement of the crystal structure of gallium, Zeitschrift für Kristallographie-Crystalline Materials. 117 (1962) 293-300.

[73] D. Schiferl, 50‐kilobar gasketed diamond anvil cell for single‐crystal x‐ray diffractometer use with the crystal structure of Sb up to 26 kilobars as a test problem, Rev. Sci. Instrum. 48 (1977) 24-30.

[74] R. Blunt, W. Hosler, H. Frederikse, Electrical and Optical Properties of Intermetallic Compounds. II. Gallium Antimonide, Physical Review. 96 (1954) 576.

[75] R. Breckenridge, R. Blunt, W. Hosler, H. Frederikse, J. Becker, W. Oshinsky, Electrical and optical properties of intermetallic compounds. I. Indium antimonide, Physical Review. 96 (1954) 571.

[76] M. Leszczynski, H. Teisseyre, T. Suski, I. Grzegory, M. Bockowski, J. Jun, S. Porowski, K. Pakula, J. Baranowski, C. Foxon, Lattice parameters of gallium nitride, Appl. Phys. Lett. 69 (1996) 73-75.

[77] D.R. Lide, CRC handbook of chemistry and physics, CRC press, 2004.

[78] A. Zoroddu, F. Bernardini, P. Ruggerone, V. Fiorentini, First-principles prediction of structure, energetics, formation enthalpy, elastic constants, polarization, and piezoelectric constants of AlN, GaN, and InN: Comparison of local and gradient-corrected density-functional theory, Physical Review B. 64 (2001) 045208.

[79] C. Chuo, C. Lee, J. Chyi, Interdiffusion of In and Ga in InGaN/GaN multiple quantum wells, Appl. Phys. Lett. 78 (2001) 314.

[80] B. Ihsan, P. Gregor (Eds.), Thermochemical data of pure substances, 3rd ed., Weinheim ; New York : VCH, 1995.

[81] G. Henkelman, B.P. Uberuaga, H. Jónsson, A climbing image nudged elastic band method for finding saddle points and minimum energy paths, J. Chem. Phys. 113 (2000) 9901-9904.